\documentstyle[twocolumn,psfig,prb,aps]{revtex}
\begin{document}
\draft
\twocolumn[\hsize\textwidth\columnwidth\hsize\csname @twocolumnfalse\endcsname

\title{Period doubling in the 2D Antiferromagnet. New gauge-fields and
their anomalies.}

\author{A.G. Abanov, J.C. Talstra}
\address{James Franck Institute of the University of Chicago,
5640 S.Ellis Avenue, Chicago, IL 60637, USA}
\date{\today}
\maketitle

\begin{abstract}
We construct an effective gauge field theory model describing a 2D
quantum antiferromagnet in the flux phase ground state. Due to period
doubling the number of gauge fields is {\em four} rather than just one.
These additional gauge fields correspond
to field configurations in which the flux is staggered, which should
be taken into account to enforce single occupancy within the magnetic
unit cell.  We calculate a new
type of quantum anomaly, present in such a model.  In
particular this leads to a generalized local flux hypothesis, which
is tested numerically.\\
\end{abstract}
]

Although there is  considerable evidence to believe that the ground
state of the two dimensional spin-$\frac{1}{2}$ Heisenberg
antiferromagnet on a square lattice with nearest neighbor interactions
has N\'{e}el order, the ground
state of the doped antiferromagnet
is still very far from being understood.  We believe that one of the
possible universality classes is correctly described by the so-called
flux phase which is characterized by non-zero chirality of
spins. Motivated by this, one
can apply the mean field approach, first proposed in \cite{Anderson},
to the 2D Heisenberg model.
\begin{equation}
H=\sum_{\langle i,j\rangle} J{\bf S}_i\cdot{\bf S}_j 
\label{HeisHam}
\end{equation}
This yields the following effective Hamiltonian:
\begin{equation}
\label{model}
H=-\sum_{\langle i,j\rangle,\sigma} \Delta_{ij}c_{i\sigma}^\dagger
c_{j\sigma} + \frac{2}{J}\left| \Delta_{ij}\right| ^2
\end{equation}
Here $\Delta_{ij}$ is to be found self-consistently in such a
way that it minimizes the energy of the system. To exclude the extra
non-physical degrees of freedom one has to impose an additional
constraint
\begin{equation}
\label{constraint}
 \sum_{\sigma}c_{i\sigma}^\dagger c_{i\sigma}=1
\end{equation}
There are several known solutions for $\Delta_{ij}$ which give local
minima of the energy \cite{AffleckMarston,IoffeLarkin}. The real
question,
however, is which of them will be preferred by doping. One possible
candidate is \cite{AffleckMarston} $\Delta_{ij}= |\Delta |
e^{iA_{ij}}$, $\prod_{\rm plaquette}e^{iA_{ij}}=-1$.
This state is called the {\em flux phase} because the effective
Hamiltonian (\ref{model})---without constraint
(\ref{constraint})---describes fermions in a background magnetic
field, with flux $\pi$ (in units where the elementary flux quantum
$\frac{h}{e}$ is $2\pi$) per plaquette. We will focus on this state in
the present paper.
One can easily find the spectrum of such fermions to be:
$\epsilon(\vec{k})=\pm 2\Delta \sqrt{\cos^2k_x+\cos^2k_y} $.  The
filling factor---the ratio of the total number of electrons with given
spin to the number of lattice sites---is 1/2. This means that in the
mean field approximation the lower branch of this spectrum is
completely filled with electrons of each spin. This mean field
solution is not a very good approximation to (\ref{HeisHam}) because
it does not take into account the local constraint (\ref{constraint}).
This is why in this mean field the antiferromagnet looks like a metal
while after taking into account the constraint it becomes an
insulator. The way to deal with the constraint is to consider the
fluctuations around the saddle-point mean field with $\pi$-flux
\cite{IoffeLarkin,BaskaranAnderson,WenWilczekZee,WiegKhvesh1,Laughlin}.
More precisely these are fluctuations of the phase of $\Delta$, or
$A_{ij}$, that restore the single occupancy constraint
(\ref{constraint}). One can easily see this, if one notices that the
fluctuating gauge field forces the current to be zero on {\em every}
lattice bond. In what follows we restrict ourselves to phase
fluctuations and adopt the common point of view that fluctuations of
the modulus of Hubbard-Stratanovich field $|\Delta|$ are not relevant
for the low energy sector of the theory because they have a gap with
size of order $J$, see eg.\ \cite{Laughlin}.

From now on we will leave spin out of the discussion, since in the
models that we will discuss, both up- and down-spin electrons are
decoupled.  Let us start from
mean field theory (\ref{model}) neglecting the local
constraint (\ref{constraint}) and fluctuations of the gauge field.  To
lift the ambiguity of the sign of the flux ($-\pi\equiv \pi \; {\rm
mod} \; 2\pi$), we allow in addition to horizontal and vertical, also
diagonal hops of {\em amplitude} $\frac{m}{2}|\Delta|$ in all four diagonal directions,
effectively distributing the flux through every plaquette over its two
constituent triangles \cite{footnote1}.
Then in the gauge $A^x_{mn}=(-1)^n \frac{\pi}{2}$, $A^y_{mn}=
\frac{\pi}{2}$, $A^{\rm NE}_{mn}=(-1)^n \frac{\pi}{2}$ and $A^{\rm
NW}_{mn} = (-1)^n \frac{\pi}{2}$, where NE and NW refer to
``north-east'' and ``north-west'' diagonal bonds, corresponding to
the $\pi$-flux per plaquette we have the following spectrum:
\begin{eqnarray}
\label{spectrum1}
\epsilon(\vec{k})=\pm 2|\Delta| \sqrt{\sin^2k_x+\sin^2k_y+m^2 \cos^2k_x \cos^2k_y}
\end{eqnarray}
Because the periodicity of the lattice is doubled, due to the presence
of half a flux quantum per plaquette, the Brillouin zone is half the
size compared to the zero magnetic field case:
$-\frac{\pi}{2}<k_x<\frac{3\pi}{2},\, |k_y| <\frac{\pi}{2}$.  As one
can see from the spectrum, at half filling, all contributions to the
low energy dynamics of the theory come from the vicinity of the
``Dirac''-points $(k_x,k_y)=(0,0), (\pi,0)$. Now we introduce a
slightly different representation which is more useful for studying low
energy dynamics. We choose a $2\times 2$ unit cell on the square
lattice, reducing the Brillouin zone further to $ |k_x|,|k_y|
<\frac{\pi}{2} $.  This choice of unit cell is convenient because in
the reduced Brillouin zone there is only {\em one} Dirac point at
${\bf k} =(0,0)$, albeit doubly degenerate. The low energy
Hamiltonian can be constructed by simple Taylor expansion in momentum
around this point.  This ``lattice adopted basis'' was proposed in
\cite{Wiegmann}. The result is the massive Dirac Hamiltonian:
\begin{equation}
\label{H0}
        {\cal H}_0=\alpha_{\mu}i\partial_{\mu}+\beta m.
\end{equation}
Here $\alpha_{\mu},\beta$ with $\mu=1,2$ are $4\times 4$ matrices
obeying the Clifford algebra. The wave function has 4 components,
corresponding to the 4 sites in the unit cell (or 2 branches $\times$
2 Dirac points). The model in its current form describes two species
of Dirac fermions, in 2+1 dimensions.

Since the unit cell is now $2\times 2$, we introduce the following
convention for decomposing an object ${\cal O}$ living on the $L\times L$
lattice, into {\em four} others living on a lattice of
cells:
\begin{eqnarray}
\lefteqn{ {\cal O}({\bf r}) \equiv \sum_{i=1}^4 \Lambda^{(i)}{\cal
O}^{(i)}({\bf r})} \protect\label{sqsymm} \\
&&={\cal O}^{(1)}({\bf r})+(-)^{x+y}{\cal O}^{(2)}({\bf r})
+(-)^x{\cal O}^{(3)}({\bf r})+(-)^y{\cal O}^{(4)}({\bf r})\nonumber
\end{eqnarray}
$\Lambda=\pm 1$ is the
generalization of the sign-function: it distributes the $+,-$ signs
within the unit cell according to one of the particular
``staggered-ness'' patterns, labeled
by $\Lambda$: $
{\renewcommand{\arraystretch}{0.6}\begin{array}{|c|c|}\hline + & + \\
\hline + & +\\ \hline \end{array}}\,$,
${\renewcommand{\arraystretch}{0.6}\begin{array}{|c|c|}\hline - & + \\
\hline + & -\\ \hline \end{array}}\,$,
${\renewcommand{\arraystretch}{0.6}\begin{array}{|c|c|}\hline - & - \\
\hline + & +\\ \hline \end{array}}\,$ and
${\renewcommand{\arraystretch}{0.6}\begin{array}{|c|c|}\hline - & + \\
\hline - & +\\ \hline \end{array}}\,$.  

The main question we are going to address in this paper is how to
incorporate fluctuations of the (lattice) gauge field into the
continuum approximation (\ref{H0}). The naive answer would be just to
apply the minimal substitution to (\ref{H0}). We argue that this is
not enough to make the theory fully gauge invariant and thus to
satisfy the no-double-occupancy constraint.  Due to period doubling
the replacement ${\bf p}\rightarrow {\bf p}-e{\bf A}$ in (\ref{H0})
describes only gauge field fluctuations which are smooth inside the
unit cell, i.e.\ of type $\Lambda^{(1)}$ in (\ref{sqsymm}).  These
particular fluctuations alone would only {\em partially} restore the
constraint (\ref{constraint}), just enforcing the total number of
electrons inside each unit cell to be 4. For the constraint to hold
locally on each site we have to include {\em staggered} gauge field
fluctuations, of types $\Lambda^{(2)},\, \Lambda^{(3)}$ and
$\Lambda^{(4)}$. These staggered fields give rise to additional fields
in the continuum theory\cite{Wiegmann,Khvesh}.  A theory that leaves those
extra fields out, fails to describe {\em intra-cell charge-transfer}
correctly.  Only the exact treatment of the single occupancy
constraint can cure the mean field theory to restore insulating
properties within the unit cell.

One quick way to find which terms we should add to Hamiltonian
(\ref{H0}), is to consider the following: on the lattice we can
perform a local gauge transformation, which might be different on each
site within the unit cell. In the continuum this means, that we must
be able to apply gauge transformations (which should leave ${\cal H}$
invariant) that not only depend on ${\bf r}$, but are also different for the
4 components of the wavefunction $\psi_i({\bf r})$, $i=1\ldots 4$. I.e.\ $\psi$ is multiplied
not by a phase, but by a unitary diagonal matrix, which is no longer
proportional to the unit matrix.  Using the previously introduced
notation, we write such a general gauge field transformation
compatible with period doubling (and smooth from cell to cell, where
the word ``cell'' refers to a block of $2\times 2$ plaquettes) as
$\psi\rightarrow \exp ( i\sum_{i=1}^4 \phi_i \delta^{(i)})\psi $.
$\delta^{(i)}$ is the $4\times 4$ matrix-representation of $\Lambda $,
so $\delta^{(i)}\equiv {\rm diag}(\Lambda^{(i)}_{00},\Lambda^{(i)}_{01}, 
\Lambda^{(i)}_{10},\Lambda^{(i)}_{11})$
e.g.\ $\delta^{(4)} = {\rm diag}(1,-1,1,-1)$.
Then the simplest, minimal way to complete Hamiltonian (\ref{H0}) to
make it gauge invariant under all 4 of these transformations is
to introduce 4 gauge fields $E,F,G$, and $K$ in the following way:
\begin{equation}
{\cal H} = \left(\sum_{\mu=1,2}e^{i{\bf M}_\mu}\alpha_\mu \partial_\mu 
e^{-i{\bf M}_\mu}\right) + 
e^{i{\bf N}}\beta m e^{-i{\bf N}},
\label{H}
\end{equation}
where ${\bf M}_\mu =F_\mu+\delta^{(2)}E_\mu+\delta^{(3)}G_\mu+
\delta^{(4)}K_\mu $, ${\bf N}=\delta^{(3)}G_2+\delta^{(4)}K_1$,
provided $E,F,G$ and $K$ gauge-transform as follows: $F_\mu\rightarrow
F_\mu +\phi_1 \ldots K_\mu\rightarrow K_\mu + \phi_4$, $\mu=1,2$.
The particular form of ${\bf N}$ is dictated by the fact that we
do not allow flux fluctuations through triangles.

One can easily see that the usual minimal substitution gauge field can
be identified as $A_1=\partial_1 F_1$, $A_2=\partial_2 F_2$. The other
fields correspond to staggered configurations of the magnetic
field on the lattice as follows:
\begin{equation}
\begin{array}{ccc}
\partial_x\partial_y (F_1-F_2) &=& +\Phi^{(1)}({\bf r})\\
(E_1-E_2) &=& - \Phi^{(2)}({\bf r})\\
\partial_x (G_1-G_2) &=& -\frac{1}{2} \Phi^{(3)}({\bf r})\\
\partial_y (K_1-K_2) &=& +\frac{1}{2} \Phi^{(4)}({\bf r})
\end{array},
\label{fluxmap}
\end{equation}
and $\Phi^{(i)}({\bf r})$ is one of the symmetry components of the flux
through plaquette ${\bf r}$ via (\ref{sqsymm}).  The important point here is
that it is not possible to expand Hamiltonian (\ref{H}) to linear (or
even any higher but finite)
order in gauge-fields other than $F$ and conserve full
gauge invariance of the theory \cite{footnote2}.

We would now like to see how the electronic density responds to fluctuations
of these new fields. 
As we shall see staggered magnetic fields cause staggered density fluctuations,
so we define the field induced staggered density $\rho^{(i)}$ as
$$\rho^{(i)}({\bf r})+\frac{1}{2}\delta_{(i),(1)}\equiv \sum_{\psi:
E<0} \psi^*_E({\bf r}) \delta^{(i)}\psi_E({\bf r}), $$
The $\psi_E$ form an orthonormal basis of eigenstates in the single particle
Hilbert space. In the groundstate all negative energy
states are occupied, as indicated in the sum. To calculate these
densities, we can use the following identity\cite{anom}:
\begin{equation}
\label{AnGen}
        \rho^{(i)}({\bf k})=\int_{-\infty}^{+\infty}\!\!\frac{dz}{2\pi}
        \!\!\int\!\! \frac{d^2p}{(2\pi)^2} {\rm Tr}\langle p|
        \delta^{(i)}\frac{-{\cal H}}{z^2+{\cal H}^2}|p-k\rangle
\end{equation}

To make further progress, we have to consider the limit where
fluctuations are small. Then, to first order in fields, we obtain
after some algebra:
\begin{eqnarray}
\label{AnDen1}
\rho^{(1)}({\bf r}) & =& -\frac{{\rm sgn}(m)}{2\pi} \Phi^{(1)}({\bf r})\\
\label{AnDen2}
\rho^{(2)}({\bf r}) & =& -\frac{1}{3}\frac{{\rm sgn}(m)}{2\pi}
\partial_x\partial_y \Phi^{(2)}({\bf r})\\
\label{AnDen3}
\rho^{(3)}({\bf r}) & =& -\frac{{\rm sgn}(m)}{4\pi} \partial_y\Phi^{(3)}({\bf r})\\
\label{AnDen4}
\rho^{(4)}({\bf r}) & =& -\frac{{\rm sgn}(m)}{4\pi} \partial_x\Phi^{(4)}({\bf r})
\end{eqnarray}
Here we kept only the zeroth order in $\frac{1}{m}$ and the lowest order in
momenta. The fact that (\ref{AnDen1})-(\ref{AnDen4}) survive when
$m\rightarrow\infty $ indicates their quantum anomalous origin.
Therefore all staggered densities are expressed in terms of derivatives of
staggered fluxes on the square lattice.  Notice the non-trivial
prefactors in the field-density relationships.  Eq.\ (\ref{AnDen1})
is widely known as a manifestation of the Chern-Simons term in the
effective action of the ordinary gauge-field: $\frac{1}{2\pi}
A_0\nabla\times A$. Relations (\ref{AnDen2}-\ref{AnDen4}) are new,
and represent the generalization of the Chern-Simons term to the other
gauge-fields. 

Now we present some numerical support for these
predictions to see how they hold up on the original lattice, with Mean
Field Hamiltonian (\ref{model}). Our simulations were carried out on
lattices of $L_x\times L_y$ sites ($L_x,L_y \leq 50$), in the presence
of periodic boundary conditions (other boundary conditions do not
change our results qualitatively). We let the electrons propagate in a
uniform background magnetic field of flux $\pi$/plaquette in addition
to a small modulation. The absolute value of the hopping amplitude is
constant and non-zero only for nearest and next-nearest neighbors. 
We apply a small, circularly symmetric perturbation to the uniform
background flux $\Phi({\bf n})=\pi$, such that the additional flux
through the plaquette ${\bf n}$ is given by: $\delta\Phi({\bf n})=\pi
f({\bf n}) \Lambda^{(i)}({\bf n})$. Here $f({\bf n})$ is chosen to be
gaussian: $\pi A \exp (-{\bf n}\cdot{\bf n}/ \xi^2)$+ const. The
constant is chosen such that the total flux through the entire lattice
is unchanged.
After diagonalizing the (single particle) Hamiltonian we fill half of the
spectrum with electrons and measure $\rho^{(i)}({\bf n})$.
Notice that $\rho^{(i)}$ is defined only one quarter of the lattice sites.

In our simulations we pick the ratio of the amplitudes of diagonal to
NN hops to lie around 0.2, making the underlying field theory
massive. The mass must be such that the gap is much smaller than the
bandwidth $m\ll 1$, or $\frac{t'}{t}\ll 1$. Since the result of the
staggered density response (\ref{AnDen1}-\ref{AnDen4}) is obtained as
an expansion in $\frac{k}{m}$ we really have a window for $m$:
$\xi^{-1}\ll \frac{t'}{t}\ll 1$. In that case $\xi\gg 1$, so that the
continuum approach is valid. At the same time, $\xi$ has to be smaller
than the lattice size $L$ to avoid finite size effects: $1\ll \xi\ll
L$. Finally we should keep the amplitude of the perturbation small
enough to remain within the realm of applicability of the linear
approximation and anomaly calculation: $\delta\Phi\ll m^2$, or
$\frac{\delta\Phi}{\Phi_0}\ll \left(\frac{t'}{t}\right)^2$ .

\begin{figure}
\vspace{1.5in}
\psfig{file=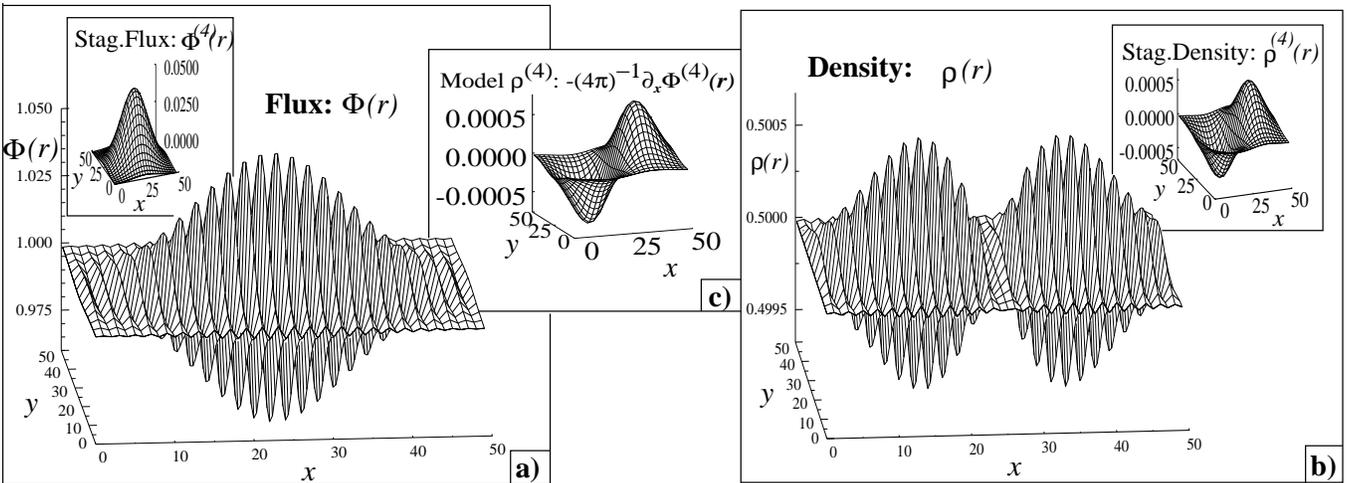,width=7in}
\widetext
\caption{Example of the application of a flux perturbation, ($A=0.05$,
$\xi=14$, $50\times 50$ sites) of type $\delta^{(4)}$.  Panel a) shows
the {\em total} applied flux, resulting from turning on field
$\Phi^{(4)}({\bf r})$ with a gaussian profile, as shown in the
inset. Panel b) shows the total density response, and the inset shows
its $\rho^{(4)}({\bf r})$ component. Panel
c) illustrates what we would expect for $\rho^{(4)}$ based on formula
(\protect\ref{AnDen4}). }
\narrowtext
\label{fluctexample}
\end{figure}

\begin{figure}
\psfig{file=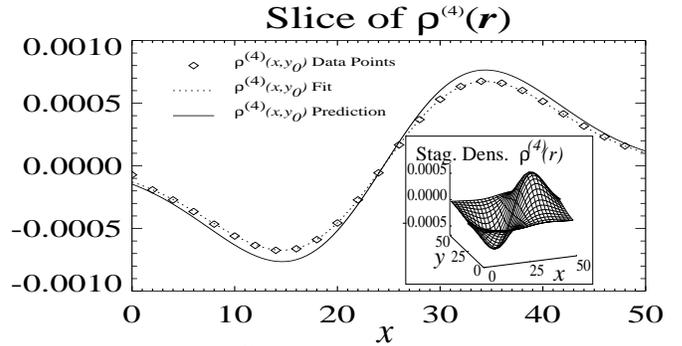,width=\hsize}
\caption{Density $\rho^{(4)}$ plotted along a line of constant $y$ for
the case of flux perturbation of type $\delta^{(4)}$ taken from the
data of fig. \protect\ref{fluctexample}. The solid line is the
predicted density response from (\protect\ref{AnDen4}) on this
slice.  There is a finite size error in the {\em amplitude} of about $10\%$,
illustrated by the dashed curve.}
\label{fluctK}
\end{figure}

\begin{figure}
\psfig{file=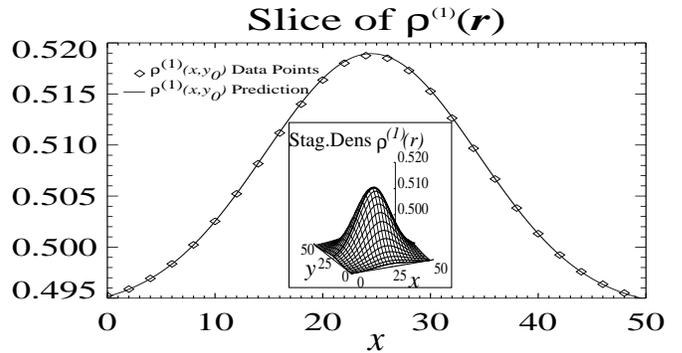,width=\hsize}
\caption{Density $\rho^{(1)}({\bf r})$ plotted along a line of
constant $y$ for the case of a flux perturbation of type $\delta^{(1)}$,
$A=0.05,\xi=14$, $50\times 50$ sites. The slice of the 2D data is
indicated by the fat line in the inset. The solid line is the predicted
density response from (\protect\ref{AnDen1}) on this slice. 
The fit is exact to within discretization errors.}
\label{fluctF}
\end{figure}

Fig.\ \ref{fluctexample} shows an example where we have chosen a gauge
fluctuation of the type $\delta^{(4)}$.  Considering the {\em a
priori} strong restrictions on the parameters to satisfy just
mentioned inequalities, the numerical experiments confirm the anomaly
predictions rather well. In fig.\ \ref{fluctK} we investigate
quantitatively the response to this perturbation. Equation
(\ref{AnDen4}) tells us that the density response should be the
derivative of a gaussian of the same width as the flux perturbation:
$\partial_x e^{-(r/\xi )^2} \sim x e^{-(r/\xi )^2} $. The data points
of fig.\ \ref{fluctK}---which show a slice through the lattice at
fixed $y=L/2$---fit this well with an amplitude that deviates $< 10\%$
from the predicted value of (\ref{AnDen4}). In fig.\ \ref{fluctF} we
repeat this exercise for the density response to the simplest gauge
fluctuation: the field $F$ (also chosen to be gaussian).  According to
(\ref{AnDen1}) the predicted response should again be gaussian, and
the data points fit this well, with an amplitude that deviates $<1\%$
from the predicted value. The $E$-field (type $\delta^{(2)}$) obeys
(\ref{AnDen2}), but for system sizes that we considered the amplitude
is $\sim 30\% $ too small.

In conclusion, we showed that to maintain the single occupancy
constraint at zero doping, extra gauge fields should be added to the
continuum field theory description of (\ref{model}). These fields
represent staggered flux fluctuations on the lattice and are relevant
for low-energy dynamics because of period doubling due to the
$\pi$-flux mean field. The gauge invariance of the lattice gauge field
theory demands that new gauge fields enter into the continuum
description (\ref{H}) as {\em non-linear} terms. The presence of these
fields leads to anomalous terms for induced {\em staggered}
densities. The situation, like the one described here, has been
encountered in gravitational anomaly calculations \cite{Witten}. In
fact our new gauge fields $E,G,K$ play the role of tetrades and
spin-connections in gravity, except for the fact that due to fermion
species doubling on the lattice, new gauge fields correspond to
generally covariant transformations which in addition mix these two
species of fermions in isospace. The new fields and new quantum
anomalies found, must be important for any calculation of quantities
which involve intra-cell charge-transfer. In fact these fluctuations
of staggered flux restore the insulating behavior of the spin system
within the unit cell.  As the fluctuations determine the holon-doublon
wavefunction, one should be able to probe them directly by high-energy
Raman scattering \cite{Raman}.

We would like to thank P.B.\ Wiegmann for collaboration during the
all stages of this project. AGA was supported by a Hulda B.\ Rotschild
Fellowship, as well as NSF Grant DMR 9509533. JCT was supported under
STCS NFS DMR 9120000.

\begin{references}
\bibitem{footnote1} 
There is no such ambiguity at non-zero doping away from half
filling. Our introduction of a mass in the Dirac spectrum through
small diagonal hopping plays the role of small doping and realizes the
parity anomaly \protect\cite{WiegKhvesh2}.
\end{references}
\end{document}